# scientific reports

# A unifying physically meaningful relativistic action

Yaakov Friedman

**The motion of an object under the influence of force fields and/or media is described by means of a world-line with least action in its influenced spacetime. For any spacetime point x and a four-vector u, measured in the frame of an inertial observer, a unifying and physically meaningful action function $L(x, u)$ generating the action is defined. To ensure independence of the observer and of the parametrization on the world-line, $L(x, u)$ must be Lorentz invariant and positive homogeneous of order 1 in u. The simplest such $L(x, u)$ depends on two four-potentials. In most cases, these potentials can be defined directly from the sources of the fields without the need for field equations. The unified dynamics equation resulting from this action, properly describes the motion in any electromagnetic field, in any static gravitational field, in a combined electromagnetic and gravitational field, as well as the propagation of light and charges in isotropic media.**

Lagrangian mechanics postulates that the motion of an object can be described by a path with *stationary* action. This action is defined by an action function - the Lagrangian function. For a stationary path, the Euler–Lagrange equations lead to Newton's second law. In this paper, we present a relativistic extension of this theory.

The stationary action principle, or principle of least action, was formulated by de Maupertuis as "Nature is thrifty in all its actions." Fermat postulated a similar principle for light propagation as "Light travels between two given points along the path of shortest time." Hilbert introduced an action for general relativity, called the Einstein–Hilbert action. Dirac demonstrated how the least action principle can be used in quantum calculations. Schwinger and Feynman independently applied this principle in quantum electrodynamics.

Despite the success of the principle of least action, however, it is not clear how to interpret actions physically. In most cases, the physics behind the definition is not clear. Moreover, in different areas of physics, the action is defined differently. This causes problems, when the motion is affected by different forces.

This paper addresses all of these issues. A simple action function which unifies the relativistic dynamics of electromagnetism, gravity and motion in isotropic media, is presented. Moreover, the physics behind our action function is clear. It is based on geometry.

Bernhard Riemann was the first to think that the laws of physics should define the geometry of space[1]. His dream was to develop the mathematics to unify the laws of electricity, magnetism, light and gravitation. His idea to replace straight lines with geodesics was used later in General Relativity (*GR*).

The application of Riemann's ideas to gravity needed two new ideas. Riemann considered how forces affect *space*, while the geometric model needs *spacetime*. Only in spacetime does Newton's First Law become geometric. Riemann used positive definite metrics. The Minkowski metric of flat spacetime, however, is not positive definite. Fifty years after Riemann's death, Einstein (in 1915) used pseudo-Riemannian geometry of spacetime as the cornerstone of *GR*, and the force of gravity was successfully modeled using geometry.

This, however, was only a partial fulfillment of Riemann's program. Since the Equivalence Principle holds only for gravitation, *GR* singles out the gravitational force from other forces which are not treated geometrically. For example, the acceleration of a charged particle in an electromagnetic field depends on its charge-to-mass ratio. Thus, the electromagnetic field does *not* create a common stage on which all particles move. Indeed, a neutral particle does not feel any electromagnetic force at all. The way spacetime curves due to an electric field depends on *both* the field *and* intrinsic properties of the object. This raises the question: Can Riemann's principle of "force equals geometry" be applied to other forces? Can Riemann's program be extended to object-dependent forces?

In this paper, several new ideas which enable us to geometrize not only gravity, but also electromagnetism and motion in media, are introduced. The model presented here is thus a continuation of Riemann's program.

By extending the Principle of Inertia to motion under the influence of force fields and media, an object will move along a world-line with least action in its influenced spacetime. The influenced spacetime is described by an action function $L(x, u)$ for any spacetime point x and a four-vector u, which are measured in the frame of an inertial observer. The action function defines the infinitesimal *distance* between two events, with coordinates x

Extended Relativity Research Center, Jerusalem College of Technology, Israel, P.O.B. 16031, 91160 Jerusalem, Israel. email: friedman@g.jct.ac.il

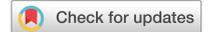

and $x + \varepsilon u$, in the object's spacetime. In order that the laws of motion will satisfy the Principle of Relativity, the action function must be Lorentz invariant. To guarantee that the laws will be independent of the choice of the evolution parameter, the action function $L(x, u)$ has to be positive homogeneous in $u$ of degree 1. Finally, if the strength of the fields tend to zero, the action function must reduce to that for empty space.

Following A. Einstein's quote, "If you can't explain it simply, you don't understand it well enough", a *simple* action function $L(x, u)$ satisfying these properties is introduced. For any electromagnetic and most gravitational fields, this action function can be derived explicitly from the sources of the field, without the need of field equations. The derivations reveal the effect of the fields on the geometry of spacetime. Applying the Euler–Lagrange equations to this action function leads to a unifying relativistic dynamics equation. Since, in relativity, the speed of a moving object is bounded by the speed of light, the acceleration, described by the evolution equation, cannot be independent of the velocity of the object. The resulting dynamic equation reveals that in relativistic dynamics, the dependence on the four-velocity is either linear or quadratic.

This dynamics equation properly describes the relativistic motion of both charged and uncharged, massive and massless particles in any electromagnetic and static gravitational field, both in vacuum and in an isotropic medium. It is shown that this model properly predicts the velocity of light in moving water (as known from the Fizeau experiment) and Snell's law for light refraction.

## Results

**Extended principle of inertia.** A geometric description of any motion may be represented by a worldline in spacetime, which depicts the position of the object at any point in time, and can be considered as the graph of the position as a function of time. To simplify the mathematical model, the motion will be represent by the use of coordinates in the frame attached to an inertial observer. These coordinates form a four-dimensional space $M$, endowed with the Minkowski metric $\eta_{\mu\nu} = diag(1, -1, -1, -1)$. We use Greek indices to label these coordinates. From Special Relativity, it is known that the real spacetime transformations between two inertial observers are the Lorentz transformations. In order that this model be consistent with the relativity principle, it should be invariant under these transformations. Thus, this theory is Lorentz covariant, as is electromagnetism, but not generally covariant, like *GR*.

The Principle of Inertia implies that the worldline of free motion is a straight line in $M$, with minimal action with respect to the Minkowski metric. Why does an object move with constant velocity or by a stationary worldline when it is not affected by forces? The answer is that an inanimate object is unable to change its velocity.

It appears that also when there are force fields and/or media which affect the object's motion, the object is still unable to change its velocity *of its own accord*. If so, then why doesn't the object appear to move along a straight line? The answer is that indeed it does not move along a straight line in the observer's flat spacetime $M$, but it does move by a "straight line" or geodesic in its influenced spacetime shaped by the forces affecting it. This is analogous to Einstein's General Relativity assumptions that (i) a gravitational field changes the spacetime geometry, and (ii) objects move in this spacetime by a stationary route. The uniqueness of the gravitational field is that it influences all objects in the same way, so one may consider spacetime as curved. But the motion of a charged particle is affected by both electromagnetic and gravitational fields, implying that its spacetime is affected by both fields. It is natural to assume that the charged particle will also move along a stationary worldline in its influenced spacetime.

This leads to formalize an extension of the Principle of Inertia in an object's influenced spacetime.

*The Extended Principle of Inertia.* Since an inanimate object (not disturbed by other objects) is unable to change its velocity, it will follow the worldline with least action in its spacetime influenced by the force fields and media affecting it.

**The action function of the influenced spacetime and its properties.** The influenced spacetime of an object is determined by the forces and the media affecting its motion and at most *one* parameter intrinsic to the object. The equivalence of gravitational and inertial mass implies that the acceleration due to a gravitational field is the same for all objects, both massive and massless. The acceleration of a charged particle moving in an electromagnetic field depends on its charge-to-mass ratio $q/m$. The motion of a photon in an isotropic medium depends on its frequency.

In order to define the shortest worldline, one has to define the "length" of a line in the influenced spacetime. For this, it is sufficient to define the distance between two infinitesimally close events $P$ and $Q$ in spacetime. This is the analog of the line element of the spacetime metric in General Relativity[2,3]. The following definition is proposed:

**Definition 1** Denote the distance in the influenced spacetime between two events with coordinates $x$ and $y$ by $D(x, y)$. The action function $L(x, u)$ for any $x, u \in M$ is defined as $L(x, u) = \lim_{\varepsilon \to 0} \frac{D(x, x + u\varepsilon)}{\varepsilon}$. This means that the distance between two events with coordinates $x$ and $x + u\varepsilon$ in influenced spacetime is $L(x, u)\varepsilon$ when $\varepsilon$ is small. The length or action of a worldline $x(\sigma)$ parameterized by an arbitrary parameter $\sigma$ on an interval $I = \{\sigma : \sigma_1 \leq \sigma \leq \sigma_2\}$ in influenced spacetime is given by

$$S[x(\sigma)] = \int_I L\left(x(\sigma), \frac{dx}{d\sigma}\right) d\sigma. \quad (1)$$

The worldline $x(\sigma)$ is called *stationary* if

$$\frac{d}{d\varepsilon}\bigg|_{\varepsilon=0} S[x(\sigma) + \varepsilon h(\sigma)] = 0 \tag{2}$$

for any smooth $h : I \to h(\sigma)$ satisfying $h(\sigma_1) = h(\sigma_2) = 0$. For motion along a worldline $x(\sigma)$, we introduce the *four-momentum*

$$p_\mu(\sigma) = \frac{\partial L(x, u)}{\partial u^\mu}\bigg|_{x=x(\sigma), u=dx(\sigma)/d\sigma}. \tag{3}$$

A worldline is stationary if for any index $\mu$, the Euler–Lagrange equations

$$\frac{d}{d\sigma} p_\mu(\sigma) = \frac{\partial L(x, u)}{\partial x^\mu}\bigg|_{x=x(\sigma), u=dx(\sigma)/d\sigma} \tag{4}$$

are satisfied.

From its definition and physical meaning, the action function $L(x, u)$ must satisfy the following properties:

1. $L(x, u)$ is scalar valued, and, to satisfy the Principle of Relativity, must be Lorentz invariant.
2. Since the action $S[x(\sigma)]$, defined by (1), must be independent of the parametrization $\sigma$, $L(x, u)$ has to be positive homogeneous of degree 1 in $u$, meaning that $L(x, au) = aL(x, u)$ for any scalar $a > 0$.
3. When the field strength approaches zero, the function becomes $L(x, u) = \sqrt{\eta_{\mu\nu} u^\mu u^\nu}$, the action in Minkowski (flat) spacetime.
4. Besides the dependence of $L(x, u)$ on the fields and media, it also depends on $k = q/m$ for a charge in an electromagnetic field, and on the frequency of a photon propagating in a medium.

**Simple action function and unifying relativistic equation of motion.** What is the simplest action function $L(x, u)$ with the required properties? Occam's razor: "explanations that posit fewer entities, or fewer kinds of entities, are to be preferred to explanations that posit more" is adopted. A. Einstein himself wrote that a "physical theory should be as simple as possible, but not simpler."

A simple way to construct a Lorentz-invariant scalar function (property 1) of variables $(x, u)$ is to contract $m$ copies of a rank one tensor $u = u^\mu$ with a tensor with $m$ lower indices. For $m = 1, 2, 3$, such functions are of the form $a_\mu(x)u^\mu$, $g_{\mu\nu}(x)u^\mu u^\nu$ and $b_{\mu\nu\eta}(x)u^\mu u^\nu u^\eta$, respectively. Also, any scalar function of these basic Lorentz-invariant scalar functions satisfy property 1. Note that, for example, $g_{\mu\nu}(x) au^\mu au^\nu = a^2 g_{\mu\nu}(x) u^\mu u^\nu$ is not homogenous of degree 1. However, $\sqrt{g_{\mu\nu}(x) u^\mu u^\nu}$ is positive homogenous of degree 1, and any linear combination of $a_\mu(x) u^\mu$, $\sqrt{g_{\mu\nu}(x) u^\mu u^\nu}$, $\sqrt[3]{b_{\mu\nu\eta}(x) u^\mu u^\nu u^\eta}$ satisfies both properties 1 and 2.

From property 3, it follows that $L(x, u)$ must contain a term of order $m = 2$, which can be written as $\sqrt{(\eta_{\mu\nu} + h_{\mu\nu}(x))u^\mu u^\nu}$, where $h_{\mu\nu}(x)$ is a symmetric tensor tending to zero when the strength of the field approaches zero.

At this point, a further simplification is proposed and the symmetric tensor $h_{\mu\nu}(x)$ is replaced with a bilinear form generated by a four-covector-valued (a rank (0, 1) tensor) function

$$h_{\mu\nu}(x) = -l_\mu(x) l_\nu(x), \quad g_{\mu\nu}(x) = \eta_{\mu\nu} + h_{\mu\nu}(x) = \eta_{\mu\nu} - l_\mu(x) l_\nu(x). \tag{5}$$

Such a metric was considered by Whitehead[4], Petrov[5], Kerr and Schild[6] and others. Hence, the term of order $m = 2$ becomes $\sqrt{\eta_{\mu\nu} u^\mu u^\nu - (l_\mu(x) u^\mu)^2}$. The minus sign is needed to avoid superluminal motion. Since the influenced direction of a gravitational field at each point in spacetime is represented by a single null covector, we assume that $l_\mu(x)$ is a null covector in the direction of the propagation of the field. This reduces the 10 free parameters of $g_{\mu\nu}(x)$ to 3 free parameters of $l_\mu(x)$. This simplification of the description of a gravitational field is based on the following conjecture.

**Conjecture 1** The action function of any gravitational field is of the form

$$L(x, u) = \sqrt{\eta_{\mu\nu} u^\mu u^\nu - (l_\mu(x) u^\mu)^2}$$

for some null covector function $l_\mu(x)$.

As shown later, the action function of the gravitational field of a spherically symmetric non-rotating body, as well as that of any static field, and of a rotating black hole, satisfy this conjecture. From property 3, it follows the coefficient of the term of order $m = 2$ is 1.

For simplicity, assume that the action function consists of terms of order 1 and 2 only. This leads to the following *simple unifying action function* for a moving object:

$$L(x, u) = \sqrt{\eta_{\mu\nu} u^\mu u^\nu - (l_\mu(x) u^\mu)^2} + k A_\mu(x) u^\mu \tag{6}$$

defined by two Lorentz-covariant, covector-valued functions $A_\mu(x)$ and $l_\mu(x)$ and a constant $k$.

The four-momentum (formula (3)) for this action function is

$$p_\lambda = \left.\frac{\partial L(x,u)}{\partial u^\lambda}\right|_{x=x(\sigma), u=dx(\sigma)/d\sigma} = \frac{\eta_{\lambda\mu}\frac{dx^\mu}{d\sigma} - l_\mu \frac{dx^\mu}{d\sigma} l_\lambda}{\sqrt{\eta_{\mu\nu}\frac{dx^\mu}{d\sigma}\frac{dx^\nu}{d\sigma} - \left(l_\mu\frac{dx^\mu}{d\sigma}\right)^2}} + kA_\lambda. \quad (7)$$

Since the action is independent of the choice of the evolution parameter, in order to simplify the formula for $p_\lambda$, a new parameter $\tau$ on the worldline is introduced by

$$d\tau^2 = \eta_{\mu\nu}dx^\mu dx^\nu - (l_\mu dx^\mu)^2, \quad (8)$$

which makes the expression in the denominator of (7) equal to 1. Note that this parameter depends on $l_\mu$, which, as shown later, describes the influence of gravitation. The parameter $\tau$ does *not* depend on $A_\mu$, describing the electromagnetic field. Since gravity affects everything, it is natural to assume that clocks will be affected by gravity. This was predicted by GR and verified experimentally by Pound and Rebka and others. On the other hand, if the clock is uncharged, it is not affected by the electromagnetic field, and if it is charged, the effect on it depends on its charge-to-mass ratio. Thus, there is no way to define the effect of electromagnetic field on time.

Denote differentiation by $\tau$ with a dot: $\dot{x} = \frac{dx}{d\tau}$. Choosing $\sigma = \tau$, the four-momentum with respect to $\tau$ is

$$p_\lambda = \eta_{\lambda\mu}\dot{x}^\mu - l_\mu \dot{x}^\mu l_\lambda + kA_\lambda. \quad (9)$$

The Euler–Lagrange Eq. (4) for a stationary worldline yield the unifying relativistic dynamics equation

$$\ddot{x}_\lambda = (l_{\mu,\nu}l_\lambda + l_{\lambda,\mu}l_\nu - l_{\nu,\lambda}l_\mu)\dot{x}^\nu \dot{x}^\mu + k(A_{\nu,\lambda} - A_{\lambda,\nu})\dot{x}^\nu. \quad (10)$$

The covector-valued function $A_\mu(x)$ will be called the *linear four-potential*, since this term in the acceleration is linear in the four-velocity. Similarly, $l_\mu(x)$ will be called the *quadratic four-potential* since the acceleration defined by it is quadratic in the four-velocity.

For a covector-valued function $f(x)$, define the rank 2 covector first order derivative $F$ by

$$F_{\lambda\nu}(f(x)) = f_{\nu,\lambda} - f_{\lambda,\nu}. \quad (11)$$

Similarly, for such a function $f(x)$, also introduce a rank 3 covector first order derivative $G$:

$$G_{\lambda\mu\nu}(f(x)) = f_{\lambda,\mu}f_\nu - f_{\nu,\lambda}f_\mu + f_{\mu,\nu}f_\lambda. \quad (12)$$

Note that the three terms of the right hand side can be obtained from the first term (as in (11)) by cycling the indices and alternating the signs. Using this notation, the *unifying relativistic equation of motion* (10) can be rewritten as

$$\ddot{x}_\lambda = G_{\lambda\mu\nu}(l(x))\dot{x}^\mu \dot{x}^\nu + kF_{\lambda\nu}(A(x))\dot{x}^\nu. \quad (13)$$

This is the relativistic analog of Newton's Second Law, which reveals that the acceleration in relativistic dynamics depends linearly or quadratically on the four-velocity of the object.

Since both four-potentials $A_\mu(x)$ and $l_\mu(x)$ are Lorentz-covariant covector-valued functions, for a field generated by a single source one can identify the form of these functions, as follows. Assume that the source is moving by a worldline $\check{x}(\tau)$ and our object is positioned at a spacetime point $P$ with coordinates $x$. Denote by $Q$ the unique intersection of the worldline of the source and the backward light cone with vertex at $P$. The relative position null four-vector $QP$ is denoted $r(x) = x - \check{x}(\tau(x))$. The time $\tau(x)$ is called the retarded time. Only the position of the source at the retarded time has influence at the spacetime position $x$, and $r(x)$ is the direction of propagation at $x$. Denote by $w(\tau(x))$ the four-velocity at the retarded time. See Fig. 1.

One may consider the acceleration and higher derivatives of the source at the retarded time, but, as shown later, it is enough to use only the four-velocity in order to define the four-potentials. Since $w(x)$ and $r(x)$ are four-vectors, one may assume that a Lorentz-covariant, four-vector valued function $f(x)$ is a multiple of these vectors by some Lorentz-invariant scalar. This multiple can be a scalar function of the Lorentz-invariant scalar product $r \circ w$ of these two vectors. Thus, Lorentz-covariant four-vectors describing the influence of a single source field at the spacetime point $x$ are of the form

$$f(x) = h_1(r(x) \circ w(\tau(x)))w(x) \quad (14)$$

or

$$f(x) = h_2(r(x) \circ w(\tau(x)))r(x) \quad (15)$$

for some scalar functions $h_1$ and $h_2$. For any field, the function $h_j$ will be defined from the Newtonian limit.

**Motion of objects in an electromagnetic field.** To identify the physical meaning of the covector function $A_\mu(x)$, consider first the motion in influenced spacetime when $l(x) = 0$. The equations of motion (13) coincide with the equation of motion of a charged particle under the Lorentz force of an electromagnetic field defined by an anti-symmetric $4 \times 4$ tensor

$$F_{\mu\nu} = A_{\nu,\mu} - A_{\mu,\nu} \quad (16)$$

and $k = q/c^2 m$.

www.nature.com/scientificreports/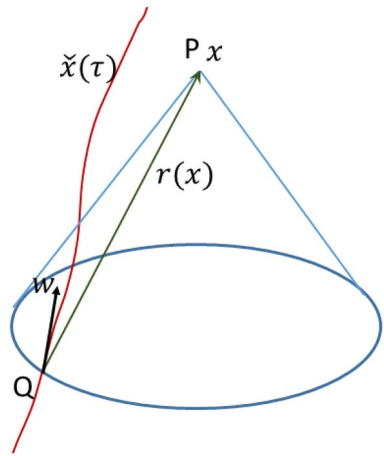

**Figure 1.** Two Lorentz covariant vector valued functions associated with a field of a single source moving along the world-line $\check{x}(\tau)$. The point $Q = \check{x}(\tau(x))$ is the intersection of the world-line with the backward light cone with vertex at $P = x$. The relative position null four-vector $QP$ is $r(x) = x - \check{x}(\tau(x))$, and $w(\tau(x))$ is the four-velocity at the retarded time.

The covector field $A_\mu$ is the linear four-potential of the field describing the geometry of the spacetime influenced by the electromagnetic field. This four-potential can be obtained by integrating the four-potentials of single sources generating the field. To define the single source field potential, we use a Lorentz-covariant four-vector of form (14). Comparing the classical acceleration of a charge with charge-to-mass ratio $q/m$ in a field of a charge $Q$ positioned at the origin with the one given by (10) and use of (14), one obtains that $h_1(y) = y^{-1}$ and the linear four-potential, denoted by $A^s_\mu(x)$ (label $s$ for single source), of a field generated by a single charge $Q$ is

$$A^s_\mu(x) = \frac{Q w_\mu(x)}{r(x) \cdot w(x)}, \qquad (17)$$

where $r(x) = r_\mu(x)$ is the relative position from the source at the retarded time, and $w(x)$ is the four-velocity at the retarded time. This is the known Liénard–Wiechert potential of a single source field[7].

The derivatives of the vectors $r(x)$ and $w(x)$ are

$$r_{\nu,\mu} = \eta_{\nu\mu} - \frac{w_\nu r_\mu}{r \cdot w}, \quad w_{\nu,\mu} = \frac{a_\nu r_\mu}{r \cdot w}, \qquad (18)$$

where $a$ is the four-acceleration at the retarded time. Using these formulas and $A^s_\mu(x)$, one obtains the field tensor (16) of a moving charge (see[8]). This field decomposes into the Coulomb field, which falls off at large distances like the square of the distance, and the radiation field, which depends on the acceleration and falls off at large distances like the distance.

One might try to use (15) to obtain the four-potential a field of a charge $Q$ positioned at the origin. For this case, one obtains that $h_2(y) = y^{-2}$ for the electromagnetic field. Thus, the linear four-potential, denoted by $\tilde{A}^s_\mu(x)$ is

$$\tilde{A}^s_\mu(x) = \frac{Q r_\mu(x)}{(r(x) \cdot w(x))^2}. \qquad (19)$$

When calculating the field tensor (16) of a moving charge based on $\tilde{A}^s_\mu(x)$ and (18), one obtains the same Coulomb field as predicted using $A^s_\mu(x)$ (Eq. (17)). However, there is no radiation field predicted. Since the radiation fields do exist, the four-potential $\tilde{A}^s_\mu(x)$ cannot be used to describe the geometry of the electromagnetic field of a single source.

To understand the influence of the electromagnetic field on the geometry of the spacetime, consider the influence of the field generated by a charge $Q$, at rest at the origin, on a charge with charge-to-mass ratio $q/m$. Using (6) and (17), the action $L(x, u)$ at a point $(x, \mathbf{x})$ is

$$L(x, u) = \sqrt{(u^0)^2 - (u^1)^2 - (u^2)^2 - (u^3)^2} - \frac{qQ}{mc^2 x} u^0.$$

From the geometric meaning of the action function $L(x, u)$ as the infinitesimal distance in influenced spacetime between $x$ and $x + \varepsilon u$, it is apparent that for a given $u^0$, the scaling of the distance for all spatial displacements with respect to empty space is the same. This implies that the geometry of this spacetime is conformal. It is known that the geometry associated to an electromagnetic field is conformal.

The sources of any electromagnetic field are currents. By the linearity of the electromagnetic field, the four-potential $A_\mu(x)$ of a field generated by currents is obtained by integrating the four-potentials $A^s_\mu(x)$ over the backward light cone with vertex at $x$. This is a way to obtain the electromagnetic field description without using Maxwell's field equations.

The classical formula for the gravitational acceleration between two masses is similar to the acceleration between two charges of opposite sign. So why can't one use the linear four-potential to describe the gravitational field?

To understand this, consider the motion of a charge with charge-to-mass ratio $q/m$ in a field generated by a single charge $Q$ positioned at the origin of the spatial axes. Let the position of the moving charged particle at some time $t$ be $\mathbf{x} \in R^3$. Then the relative position four-vector is $r = (x, \mathbf{x})$. Since $w = (1, 0, 0, 0)$, at any space point $\mathbf{x} \in R^3$, the non-zero components of the tensor $F_c$ defined by (16) and (17) are

$$F_{0j} = \frac{Qx_j}{x^3}, \quad F_{j0} = -\frac{Qx_j}{x^3},$$

for $j = 1, 2, 3$. Using (10), this implies that $\ddot{x}_j = -k\frac{Qx_j}{x^3}$, and the 3D acceleration is in the radial direction. Thus, if $\Pi$ denotes the plane in $R^3$ generated by the position vector $\mathbf{x}(0)$ of the moving particle and its velocity $\mathbf{v}(0)$ at the initial time, the acceleration will also be in $\Pi$. This implies that the motion of the particle will remain in $\Pi$ for all times. By rotating the axes, we assume that $\Pi$ is the $x^1, x^2$ plane, or $x^3 = 0$.

Use polar coordinates $\rho, \varphi$ in the plane $x^1, x^2$ and $\varrho, \phi$ in the plane $u^1, u^2$, and $\tau$ defined by (8). In these coordinates, $kA_0^s = \frac{qQ}{mc^2\rho} = \frac{\alpha}{\rho}$ and $kA_\varphi^s = 0$. Since the action function $L(x, u)$, defined by (6), depends only on $\rho$ and is independent of $t$ and $\varphi$, the four-momentum components $p_0$ and $p_2$, defined by (9), are conserved:

$$p_0 = c\dot{t} - \frac{\alpha}{\rho} = \beta, \quad p_\varphi = \rho^2 \dot{\varphi} = l \tag{20}$$

for some constants $\beta, l$. From (8), the square of length of the four-velocity in our coordinates is

$$c^2\dot{t}^2 - \dot{\rho}^2 - \rho^2\dot{\varphi}^2 = 1. \tag{21}$$

Substituting in this equation the expressions for $\dot{t}$ and $\dot{\varphi}$ from (20), yields

$$\dot{\rho}^2 + \frac{l^2 - \alpha^2}{\rho^2} - \frac{2\alpha\beta}{\rho} = \beta^2 - 1. \tag{22}$$

To describe the trajectory, introduce a function $f(\varphi) = \frac{1}{\rho(\varphi)}$, which is proportional to the classical potential energy on the trajectory. Equation (22) becomes

$$l^2(f')^2 + (l^2 - \alpha^2)f^2 - 2\alpha\beta f = \beta^2 - 1.$$

Differentiating this equation and dividing by $2l^2 f'$ yields

$$f'' + \left(1 - \frac{\alpha^2}{l^2}\right)f = \frac{\alpha\beta}{l^2}. \tag{23}$$

The solution of this equation is

$$f(\varphi) = \frac{\alpha\beta}{l^2 - \alpha^2} + A\cos\sqrt{1 - \frac{\alpha^2}{l^2}}(\varphi - \varphi_0). \tag{24}$$

The trajectory $\rho(\varphi)$ of the particle is thus a precessing ellipse with precession $\pi\alpha^2/l^2$ per revolution.

If gravity could be described by the linear four-potential, the same analysis could be applied for the motion of Mercury around the Sun. In this case $\alpha = r_s$, the Schwarzschild radius of the Sun. It is known that Mercury's orbit is a precessing ellipse, but the observed precession differs from that predicted by (24). Thus, gravitation cannot be described by a linear four-potential, but, as shown next, it is described properly by the quadratic four-potential.

**Motion of objects in a gravitational field.** Let us assume that the relativistic description of the gravitational field satisfies the Newtonian limit. This means that for any point $P$ in spacetime, the acceleration, predicted by (10), of an uncharged object ($k = 0$) positioned at $P$ with zero initial velocity coincides with the Newtonian acceleration. From (8), the four-velocity of an object at rest is $\dot{x}_r = \alpha(1, 0, 0, 0)$, with $\alpha = 1/\sqrt{1 - l_0^2}$. Since $l_\mu \dot{x}_r^\mu = \alpha l_0, l_{\mu,\lambda} \dot{x}_r^\mu = \alpha l_{0,\lambda}$ and $l_{\lambda,\mu} \dot{x}_r^\mu = \alpha l_{\lambda,0}$, the acceleration, defined by (10) of an object at rest is

$$\ddot{x}_\lambda = \frac{1}{1 - l_0^2}(l_{\lambda,0}l_0 - l_{0,\lambda}l_0 + l_{0,0}l_\lambda). \tag{25}$$

For any static gravitational field, $l_{\lambda,0} = 0$. Hence, the 3D acceleration of an object at rest in such a field is $\frac{d^2\mathbf{x}}{dt^2} = -\frac{c^2}{2}\nabla l_0^2$. Comparing this with the Newtonian acceleration $\nabla \Phi$ (lowering the index of acceleration eliminated the usual minus sign), where $\Phi$ denotes the Newtonian potential of the field, yields $-\frac{c^2}{2}\nabla l_0^2 = \nabla \Phi$. Using the fact that both $l_0^2$ and $\Phi$ must vanish at infinity, one obtains

$$l_0^2(\mathbf{x}) = -\frac{2}{c^2}\Phi(\mathbf{x}) := \phi(\mathbf{x}), \tag{26}$$

where $\phi(\mathbf{x})$ denotes the unit free, positive Newtonian gravitational potential.

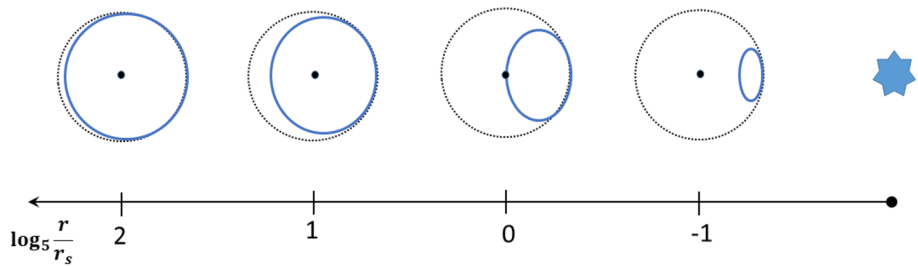

**Figure 2.** The geometry of the influenced spacetime in the vicinity of a Black Hole. At each distance, we see the 2D section of the light cone at $u^0 = 1$ in flat spacetime (black) and influenced spacetime (blue).

The null covector $l$ can be written as $l = l_0(1, \mathbf{n})$, where $\mathbf{n}$ is a norm one vector in $\mathbb{R}^3$ in the direction of the propagation of the field. Since the propagation of a static gravitational field with potential $\Phi(\mathbf{x})$ is in the direction of $\nabla \Phi(x)$, by the definition of $l$, $\mathbf{n} = \nabla \phi(x)/|\nabla \phi(x)|$. Thus, the action function for a static gravitational field with Newtonian potential $\Phi(\mathbf{x})$ is

$$L(x, u) = \sqrt{(u)^2 - (l_\mu(x)u^\mu)^2}, \quad l(x) = \sqrt{\phi(x)}(1, \nabla \phi(x)/|\nabla \phi(x)|), \tag{27}$$

with $\phi(x)$ defined by (26). It is well known[3] that the gravitational potential $\Phi(x)$ is determined by the Poisson equation and depends linearly on the sources of the field.

Consider the gravitational field outside of a static, spherically symmetric, non-rotating body of mass $M$. Place the origin of the spatial axes at the center of the body. For an object positioned at rest at a space point $\mathbf{x}$, the relative position is $r = (x, \mathbf{x})$ and $r \cdot w = x$. The classical potential $\Phi(x) = -\frac{GM}{x}$, implying that $\phi(x) = \frac{2GM}{c^2 x} = \frac{r_s}{x}$, where $r_s = \frac{2GM}{c^2}$ is the Schwarzschild radius. Thus, (27) can be written as

$$L(x, u) = \sqrt{(u)^2 - (l_\mu(x)u^\mu)^2}, \quad l(x) = \sqrt{\frac{r_s}{x^3}}(x, -\mathbf{x}). \tag{28}$$

This corresponds to (15) for the quadratic four-potential of a single source. As shown later, this four-potential predicts exactly the anomalous precession of Mercury's orbit, while a four-potential of the form (14) leads to an incorrect precession.

If a gravitational field is generated by several spherically symmetric, non-rotating bodies, then the Newtonian limit implies that the acceleration of the field is the sum of accelerations due to each source. This is equivalent to the assumption that the rank 3 tensor $G$ in the equation of motion (13) is the sum of the rank 3 tensors of each source.

To understand the change in the geometry due to gravitation, compare the 2D section of the light cone for $u^0 = 1$ in flat spacetime with spacetime influenced by the gravitational field of a Black Hole, see Fig. 2. Observe that the Black Hole pulls the light cone toward itself, but leaves it inside the cone of flat spacetime. Thus, even in presence of a strong gravitation field, the speed of a moving object, as observed by an inertial observer, will be bounded by the speed of light.

We show now that the motion in a gravitational field of a spherically symmetric body defined by the action function (28) passes the tests of *GR*. Consider the motion of an object in gravitational field of a spherically symmetric mass $M$ centered at the origin of our inertial frame. As shown above, the 3D acceleration of the object is in the radial direction. Thus, if $\Pi$ is the plane in $R^3$ generated by the position vector $\mathbf{x}(0)$ of the moving particle and its velocity $\mathbf{v}(0)$ at the initial time, the acceleration will also be in $\Pi$. This implies that the motion of the particle will remain in $\Pi$ for all times. By rotating the axes, assume that $\Pi$ is the $x^1, x^2$ plane, or $x^3 = 0$.

Use polar coordinates $\rho, \varphi$ in the plane $x^1, x^2$ and $\varrho, \phi$ in the plane $u^1, u^2$, and $\tau$ defined by (8). In these coordinates, the action function is

$$L(x, u) = \sqrt{(u^0)^2 - \Delta \varrho^2 - (\varrho \Delta \phi)^2 - \frac{r_s}{\rho}(u^0 + \Delta \varrho)^2}.$$

Since this action function depends only on $\rho$ and is independent of $x^0$ and $\varphi$, the four-momentum components $p_0$ and $p_2$, defined by (9), are conserved:

$$p_0 = \left(1 - \frac{r_s}{\rho}\right)\dot{x}^0 - \frac{r_s}{\rho}\dot{\rho} = \beta, \quad p_\varphi = \rho^2 \dot{\varphi} = l \tag{29}$$

for some constants $\beta, l$. From (8), the square of length of the four-velocity in these coordinates is

$$\left(1 - \frac{r_s}{\rho}\right)(\dot{x}^0)^2 - 2\frac{r_s}{\rho}\dot{x}^0 \dot{\rho} - \left(1 + \frac{r_s}{\rho}\right)\dot{\rho}^2 - \rho^2 \dot{\varphi}^2 = 1. \tag{30}$$

Substituting in this equation the expressions for $\dot{x}^0$ and $\rho \dot{\varphi}$ from (29) and multiplying by $1 - \frac{r_s}{\rho}$, one obtains

$$\left(\beta + \frac{r_s}{\rho}\dot{\rho}\right)^2 - 2\frac{r_s}{\rho}\left(\beta + \frac{r_s}{\rho}\dot{\rho}\right)\dot{\rho} - \left(1 - \frac{r_s^2}{\rho^2}\right)\dot{\rho}^2 - \frac{l^2}{\rho^2}\left(1 - \frac{r_s}{\rho}\right) = \left(1 - \frac{r_s}{\rho}\right). \quad (31)$$

Opening the parentheses and simplifying leads to

$$\dot{\rho}^2 + \frac{l^2}{\rho^2}\left(1 - \frac{r_s}{\rho}\right) - \frac{r_s}{\rho} = \beta^2 - 1. \quad (32)$$

To describe the trajectory, introduce a function $f(\varphi) = \frac{r_s}{\rho(\varphi)}$, which is the unit-free, classical potential energy on the trajectory. Using (29), this implies that $\dot{\rho} = \frac{l}{r_s}f'$. For this function, the above equation becomes

$$\frac{l^2}{r_s^2}(f')^2 + \frac{l^2}{r_s^2}(f^2 - f^3) - f = \beta^2 - 1.$$

Differentiating this equation and dividing by $\frac{2l^2}{r_s^2}f'$ yields

$$f'' + f = \mu + 1.5f^2, \quad \mu = \frac{r_s^2}{2l^2}. \quad (33)$$

The solution[3,9] of this equation defines the trajectory $\rho(\varphi)$ of the particle as a precessing ellipse, with precession $3\pi\mu$ per revolution. This is the observed anomalous precession of Mercury's orbit and the precession for any other observed relativistic orbits.

For light propagation in such a gravitational field, one needs to replace the 1 on the right-hand side of Eq. (30) by 0. This leads to removing the $-1$ from the right-hand side of Eq. (32). The resulting equation is the basis for the derivation of light deflection and the Shapiro time delay[10] and[11]. Thus, the motion of uncharged bodies with respect to our action function passes all relativistic solar tests predicted by *GR*.

Formula (28) can be extended to the gravitational field of a spherically symmetric, non-rotating body moving with respect to the frame of the observer. Using (15), the quadratic four-potential for such a field is

$$l_\mu = \sqrt{\frac{r_s}{(r(x) \circ w(\tau(x)))^3}} r_\mu(x). \quad (34)$$

This is the gravitational field analog of the Liénard–Wiechert four-potential of a moving source. Using (18) and (10), one can calculate the acceleration generated by this four-potential. As in electromagnetism, one obtains here two fields: a near field, which falls off at large distances like the square of the distance, and a radiation field, depending on the acceleration of the source and falling off at large distances like the distance.

Since most gravitational fields are generated by a collection of moving spherically symmetric bodies, to make our theory of gravity complete, we need to prove the following conjecture:

**Conjecture 2** The rank 3 tensor *G* in the equation of motion (13) for the evolution in a gravitational field generated by a collection of moving, spherically symmetric bodies is the sum of the *G* tensors of each source. (34)

For the gravitational field generated by a binary star, this conjecture predicts gravitational waves similar to those predicted by *GR*.

As shown in[12], the approximations used in *GR*[13,14] to describe the gravitational field of a binary star reduce to that of a single spherically symmetric body, which can be described by the action function (28). This implies that motion with such an action function will pass all the relativity tests based on binaries. As shown in[15,16], the Kerr metric describing the gravitational field of a rotating black hole is also described by an action function $L(x, u) = \sqrt{(\eta_{\mu\nu} - l_\mu l_\nu)u^\mu u^\nu}$ as in (6).

For a single point, static source, our description is similar to Whitehead's theory of gravitation[4], which is equivalent to the Schwarzschild metric, used in *GR*, see[17,18]. A gravitational field generated by several objects in Whitehead's theory is described by a metric $g_{\mu\nu}(x)$ as in (5), with $h_{\mu\nu}(x)$ the sum of such terms for all objects in the source. The predictions of Whitehead's theory deviate from those of *GR*. For example, Whitehead predicts a different formula for the gravitational field of an extended spherically symmetric body, a different formula for the precession of elliptic orbits, and a different acceleration of the center of mass of binaries. In[19], Whitehead's theory was criticized based on these predictions which contradict experimental results. Our description does coincide with Whitehead's theory of gravitation, but only for a single point static source. The metric of a field of a static source (27) differs from the one obtained in Whitehead's theory. This can be seen already for a field with two sources. Our description of a gravitational field of an extended spherically symmetric body is equivalent to that of *GR*, and the precession of orbits is exactly as in *GR*. Thus, Gibbons and Will's criticism of Whitehead's theory does not apply to our approach.

For a charged particle *q* of mass *m* moving in an electromagnetic field defined by $A_\mu(x)$ and a gravitational field defined by $l_\mu(x)$, the action function is defined by (6). In the case of a static gravitational field, $l_\mu$ is defined by (27). The evolution parameter $\tau$ is defined by (8) and is independent of the electromagnetic field. Furthermore, from formula (9), we see that the contributions of the electromagnetic and gravitational fields to the four-momentum are additive. Since the dynamics in an electromagnetic field is linear in the four-velocity of the moving object and the dynamics in a gravitational field is quadratic in this four-velocity, it does not look right to describe a field of a charged massive body by one metric containing both gravity and electromagnetism.

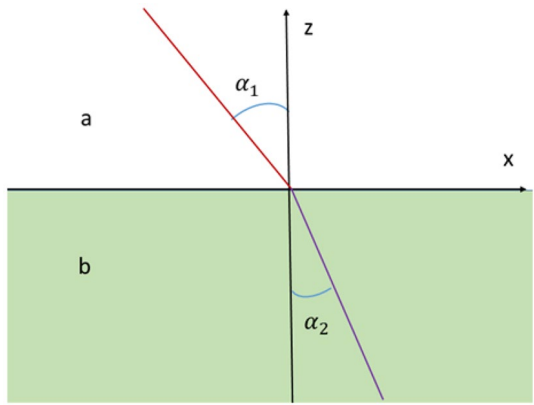

**Figure 3.** Refraction of light between two media.

**Propagation of light and charges in a medium.** Propagation of light in an isotropic medium at rest can be considered as the motion of photons in the spacetime influenced by this medium. Since the charge $q$ of the photon is zero, the parameter $k = 0$ in (6). The assumption that the medium is isotropic at rest in an inertial frame $K$ implies that $l = (l_0, 0, 0, 0)$, where the parameter $l_0$ depends only on the medium and possibly on the frequency of the photon. The dependence of $l_0$ on the frequency can be observed in the refraction of light through a glass prism, where photons with different frequencies (color) are refracted by different angles.

One knows from Special Relativity that the worldline of a massless particle, like a photon, is null. We assume that the same property holds for the motion of the photon in the influenced spacetime. This implies that $l_0^2 = 1 - \left(\frac{v_0}{c}\right)^2$, where $v_0$ is the speed of light or the speed of the photon in this medium. Using (6), the action function for the motion of a photon in an isotropic medium is

$$L(x, u) = \sqrt{\frac{1}{n^2}(u^0)^2 - (u^1)^2 - (u^2)^2 - (u^3)^2}, \qquad (35)$$

where the refractive index $n = \frac{c}{v_0}$. Thus, the metric of the photon's influenced spacetime in the medium is equivalent to the empty spacetime metric, with the speed of light in vacuum replaced by the speed of the photon in the medium.

To verify the validity of our action function for a medium, we check its prediction of the speed of light in a moving medium. By the Principle of Relativity, propagation of light in a moving medium is similar to its motion in a medium at rest, as observed by an inertial observer moving with velocity opposite to the velocity of the medium. For such an observer, the emitter of light is also moving. However, as it is known, this does not affect the speed of light. Assume that the medium moves with velocity $v_m$ in the $x$-direction, which is equivalent to observing it from a frame $K'$ moving with velocity $-v_m$ in the $x$-direction.

To describe the action function in $K'$, one needs to perform a Lorentz transformation $\Lambda$ of the co-vector $l$ from $K$ to $K'$. So, in $K'$, the action function is defined by $l' = \Lambda l = l_0 \gamma(v)(1, -\beta, 0, 0)$, with $\beta = v_m/c$. This action function defines two speeds of light in the $x$-direction

$$v_{1,2} = \frac{v_m\left(1 - \frac{1}{n^2}\right) \pm (1-\beta^2)v_0}{1 - \frac{\beta^2 v_0^2}{c^2}}.$$

Since $\beta^2 \ll 1$, ignore terms with $\beta^2$ and obtain that the velocity of light in the direction of the motion of the medium is $v_+ = v_0 + v_m\left(1 - \frac{1}{n^2}\right)$, while its velocity in the opposite direction is $v_- = -v_0 + v_m\left(1 - \frac{1}{n^2}\right)$. These results agree with the observed velocities of light in moving water in the Fizeau experiment.

Consider now the refraction of a light ray propagating between two media, separated by a plane $z = 0$, see Fig. 3. Let $n(z) = n_1$ for $z \geq 0$ and $n(z) = n_2$ for $z \leq 0$. Choose the $x$-axis in a way that the incoming ray will be in the plane $y = 0$.

Since the action function (35) is independent of $x^0, x^1, x^2$, the four-momenta $p_0, p_1 = p_x, p_2 = p_y$ on the worldline parameterized by $t$ are preserved. But $p_0 = \frac{u^0}{n^2 L} = \frac{c}{n^2 L}$ and $p_x = -\frac{u^1}{L} = \frac{v_x}{L}$, implying that $\frac{p_x}{p_0} = \frac{v_x n^2}{c}$ is preserved on a stationary worldline.

For $z \geq 0$, the $x$-component of the velocity is $v_x = \frac{c}{n_1} \sin \alpha_1$, where $\alpha_1$ is the angle of incidence for the incoming ray, and $\frac{p_x}{p_0} = n_1 \sin \alpha_1$. Similarly, $\frac{p_x}{p_0} = n_2 \sin \alpha_2$ for $z \leq 0$. Since $\frac{p_1}{p_0}$ is preserved on stationary worldlines, we obtain

$$n_1 \sin \alpha_1 = n_2 \sin \alpha_2.$$

This is Snell's law for light refraction. The derivation reveals that this law is an expression of momentum conservation of the moving photon.

The motion of a charge in an electromagnetic field with four-potential $A_\mu$ in an isotropic medium expressed by $l_\mu$, as in (35), is described by the equation of motion (13). Since, for an isotropic medium $l(x)$ is constant, the tensor $G = 0$, implying that the description of the motion of the charge is similar to that of its motion in vacuum, with the modification of the evolution parameter $\tau$ defined by (8).

## Summary and discussion

A unifying relativistic approach for motion of objects, both massive and massless, charged and uncharged, under the influence of fields and isotropic media, was presented. Riemann's principle "force equals geometry", means that a force causes space to curve and the motion of an object is along a geodesic in this curved space. This principle is the basis of *GR* (general theory of gravitation). In this paper, several new ideas which enabled the geometrization not only of gravity, but also of electromagnetism and motion in media were presented.

"Results" section introduced the notion of influenced spacetime, explained why motion should be along a geodesic, and presented a new extended principle of inertia: " Since an inanimate object (not disturbed by other objects) is unable to change its velocity, it will follow the worldline with least action in the influenced spacetime, the spacetime influenced by force fields or media affecting it".

To implement this principle, "Summary and discussion" section, Definition 1, introduced an action function $L(x, u)$ for $x$, $u$ measured in the frame of an inertial observer. This action function describes the infinitesimal distance between $x$ and $x + \varepsilon u$ in the influenced spacetime. In order that the action should be independent of the choice of an inertial observer, $L(x, u)$ should be Lorentz invariant. Moreover, in order that the action will be independent of the choice of the evolution parameter, $L(x, u)$ has to be positive homogeneous of degree 1 in $u$. Finally, if the fields tend to zero, the action function should reduces to the action function of empty space.

Based on these properties, the form of a simple action function (6) was derived in "Simple action function and unifying relativistic equation of motion". This function depends on two covector fields $A_\mu, l_\mu$ and one parameter of the moving object. Using the Euler–Lagrange equations, a unifying relativistic Eq. (13) defining the acceleration of of the moving object was derived. One component of the acceleration defined by $A(x)$ is linear in the four-velocity, and $A(x)$ was called the *linear four-potential*. Similarly, the second component of the acceleration defined by $l(x)$ is quadratic in the four-velocity and $l(x)$ was called the *quadratic four-potential*. For a field generated by a single source two forms of Lorentz-covariant four-potentials were identified.

In "Motion of objects in an electromagnetic field" it was shown that the linear four-potential $A(x)$ is a four-potential of an electromagnetic field. For a field generated by a single charge it is the known Liénard–Wiechert potential. For this field, the geometry of the influenced spacetime was found to be conformal. For a general electromagnetic field, the four-potential and the field could be obtained directly from the sources of the field, without the need for Maxwell field equations. The precession of elliptic orbits in a field of a single charge was also defined. The calculation revealed that unlike the electromagnetic field, the gravitational field can not be described by a linear four-potential.

In "Motion of objects in a gravitational field", assuming that a gravitational field satisfies the Newtonian limit, a description (27) of any static gravitational field by means of a quadratic four-potential $l(x)$ was obtained. We also found The quadratic four-potential $l(x)$ of a gravitational field generated by a moving spherically symmetric body (34) was obtained. It was shown that the model predicts gravitation radiation. Figure 2 presents the geometry of the influenced spacetime in the neighbourhood of a Black Hole. The precession of elliptical orbits, predicted by this model, was shown to coincide with the one predicted by *GR* and verified experimentally. The prediction of other GR tests by this model was also indicated. The connection of this model to the Whitehead's theory of gravitation was explained. A conjecture needed to to make this theory a self-contained theory of gravity was formulated. The way to describe the relativistic motion of charges in a combined electromagnetic and gravitation field was also described.

In "Propagation of light and charges in a medium", the propagation of light in an isotropic medium was discussed. The influence of the medium is described by a constant quadratic four-potential $l_\mu$. It was shown that this description properly predicts the speed of light in moving water as in the Fizeau experiment. The Snell's law for light refraction in this model results from the conservation of the four-momentum of the moving photon.

The dynamics presented in this paper uses the action function only to second order, and four-potentials depending only on the position and velocity of the sources. The method can also be used to study higher order derivative field theories. An example of such higher order theory for electrodynamics is Podolsky electrodynamics[20], and for gravity, see[21].

## Data availability

All data generated or analysed during this study are included in this published article.

### Acknowledgements
The author wishes to thank Menachem Steiner, Israel Felner, Tzvi Scarr and Lawrence Horwitz for helpful comments and discussions.

### Author contributions
The research presented in the paper and writing the manuscript was done by the author alone.

### Competing interests
The author declare no competing interests.

### Additional information
**Correspondence** and requests for materials should be addressed to Y.F.

**Reprints and permissions information** is available at www.nature.com/reprints.

**Publisher's note**  Springer Nature remains neutral with regard to jurisdictional claims in published maps and institutional affiliations.